\documentclass[journal=prb,manuscript=article,twocolumn,amsmath,amssymb,showpacs,showkeys,superscriptaddress]{revtex4-1}

\usepackage{graphicx}
\usepackage{amssymb}
\usepackage{bm}
\usepackage{soul}
\usepackage{subfigure}
\usepackage{graphpap}
\usepackage{amsmath}
\usepackage{xcolor}
\usepackage{hyperref}
\usepackage{multirow}
\usepackage[columnwise]{lineno}
\usepackage{enumitem}
\usepackage{natbib}
\hypersetup{
    colorlinks=true,
    linkcolor={blue!70!black},
    citecolor={blue!70!black},
    urlcolor={blue!70!black}
}
\definecolor{Blu}{rgb}{0.00,0.00,1.00}
\definecolor{Red}{rgb}{1.00,0.00,0.00}
\definecolor{Green}{rgb}{0.00,0.60,0.00}

\def\cm-1{cm$^{-1}$}
\def\cmeV{cm$^{-1}$/eV }
\def\cmeVno{cm$^{-1}$/eV}

\begin{document}

\title{Multi-wavelength Raman Spectroscopy of Ultra-narrow Nanoribbons Made by Solution-mediated Bottom-up Approach}

\author{Daniele \surname{Rizzo}}
\affiliation{School of Chemistry, Manchester University, Oxford Road, Manchester M139PL, United Kingdom}
\author{Deborah \surname{Prezzi}}
\email[]{deborah.prezzi@nano.cnr.it}
\affiliation{Nanoscience Institute of CNR, S3 Center, 41125 Modena, Italy}
\author{Alice \surname{Ruini}}
\affiliation{Nanoscience Institute of CNR, S3 Center, 41125 Modena, Italy}
\affiliation{Department of Physics, Mathematics, and Informatics, University of Modena and Reggio Emilia, 41121 Modena, Italy}
\author{Vaiva \surname{Nagyte}}
\affiliation{School of Chemistry, Manchester University, Oxford Road, Manchester M139PL, United Kingdom}
\author{Ashok \surname{Keerthi}}
\affiliation{Max Planck Institute for Polymer Research, Ackermannweg 10, 55128 Mainz, Germany}
\author{Akimitsu \surname{Narita}}
\affiliation{Max Planck Institute for Polymer Research, Ackermannweg 10, 55128 Mainz, Germany}
\author{Uliana \surname{Beser}}
\affiliation{Max Planck Institute for Polymer Research, Ackermannweg 10, 55128 Mainz, Germany}
\author{Fugui \surname{Xu}}
\affiliation{School of Chemistry and Chemical Engineering, Shanghai Jiao Tong University, 800 Dongchuan RD, Shanghai 200240, China}
\author{Yiyong \surname{Mai}}
\affiliation{School of Chemistry and Chemical Engineering, Shanghai Jiao Tong University, 800 Dongchuan RD, Shanghai 200240, China}
\author{Xinliang \surname{Feng}}
\affiliation{Department of Chemistry and Food Chemistry, Technische Universitat Dresden, Mommsenstrasse 4, 01062 Dresden, Germany}
\author{Klaus \surname{M\"{u}llen}}
\affiliation{Max Planck Institute for Polymer Research, Ackermannweg 10, 55128 Mainz, Germany}
\author{Elisa \surname{Molinari}}
\affiliation{Nanoscience Institute of CNR, S3 Center, 41125 Modena, Italy}
\affiliation{Department of Physics, Mathematics, and Informatics, University of Modena and Reggio Emilia, 41121 Modena, Italy}
\author{Cinzia \surname{Casiraghi}}
\email[]{cinzia.casiraghi@manchester.ac.uk}
\affiliation{School of Chemistry, Manchester University, Oxford Road, Manchester M139PL, United Kingdom}
\date{\today}


\begin{abstract}
Here we present a combined experimental and theoretical study of graphene nanoribbons (GNRs), where detailed multi-wavelength Raman measurements are integrated by accurate {\it ab initio} simulations. Our study covers several ultra-narrow GNRs, obtained by means of solution-based bottom-up synthetic approach, allowing to rationalize the effect of edge morphology, position and type of functional groups as well as the length on the GNR Raman spectrum. We show that the low-energy region, especially in presence of bulky functional groups is populated by several modes, and a single radial breathing-like mode cannot be identified. In the Raman optical region, we find that, except for the fully-brominated case, all GNRs functionalized at the edges with different side groups show a characteristic dispersion of the D peak (8--22 \cmeVno). This has been attributed to the internal degrees of freedom of these functional groups, which act as dispersion-activating defects. The G peak shows small to negligible dispersion in most of the cases, with larger values only in presence of poor control of the edges functionalization, exceeding the values reported for highly defected graphene.
In conclusion, we have shown that the characteristic dispersion of the G and D peaks offer further insight on the GNR structure and functionalization, by making Raman spectroscopy an important tool for the characterization of GNRs.

\end{abstract}

\pacs{73.22.Pr, 78.30.Ly, 31.15.A-, 31.15.E-, 71.15.Mb}
\keywords{Graphene nanoribbons; Raman spectroscopy; Functionalization; Density-functional theory}

\maketitle

\section{\label{sec:intro}Introduction}

\begin{figure}[!t]
\includegraphics[width=0.92\columnwidth]{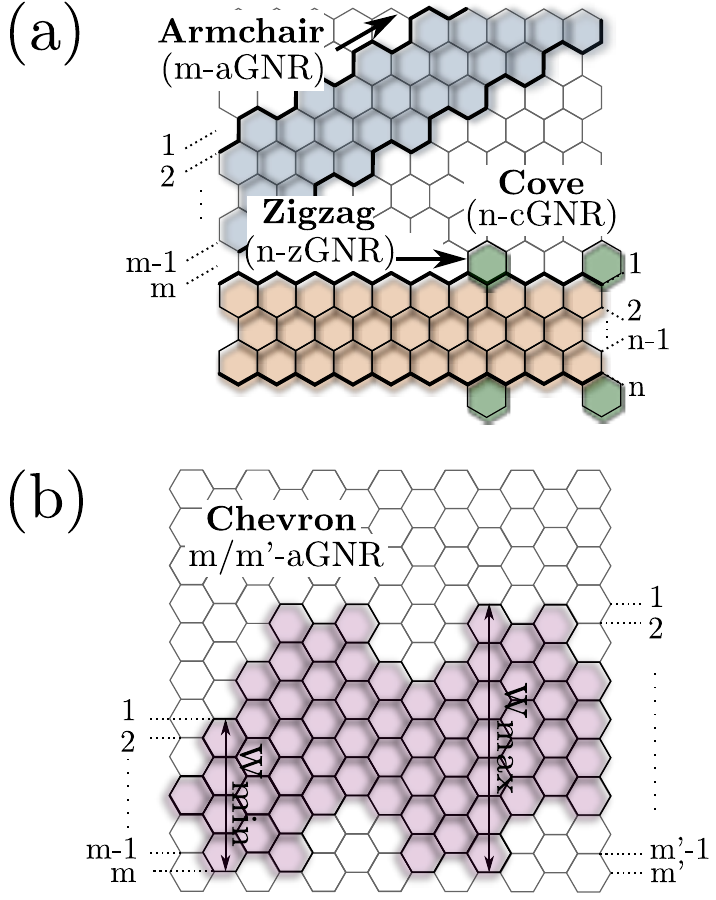}
\caption{Sketches elucidating the structure and nomenclature of GNRs with different edge geometry. Starting from standard armchair and zigzag GNRs, (a) cove-type and (b) armchair chevron-like GNRs are defined.}\label{fig:GNR_pattern}
\end{figure}

Bottom-up strategies for material fabrication, which entail a complete synthesis of a complex material starting from simple building blocks, are nowadays largely employed to produce nanomaterials and supramolecular systems with atomic-scale control \cite{shimomura-sawadaishi}. In the realm of graphene-derived systems, bottom-up approaches based on both solution-phase synthesis and surface-assisted growth have been exploited to produce structurally well-defined one-dimensional nanostructures, namely graphene nanoribbons (GNRs) \cite{energygapHan,energygap,yazyev,Gan,fujita,ezawa,brey}. These are basically  narrow strips of graphene where the outstanding properties of graphene are combined to the presence of a finite bandgap, which is derived from quantum confinement effects and makes them suitable for graphene-based electronics and opto-electronics applications \cite{FETGNR1,FETGNR2}. Besides the capability of producing ultra-narrow (width $\ll$10 nm)  and finite-gap nanoribbons with atomistic precision, solution-phase synthesis also allows for a fine tuning of their properties, that are in fact found to be highly dependent on edge morphology and functionalization \cite{yazyev,Nakada,Wakabayashi1999,sasaki,abanin,ezawa,brey,Zhihon}. Solution-based techniques allow the fabrication of GNRs with different length, edge-functionalization and edge pattern, in addition to the traditional armchair (aGNRs) and zigzag GNRs (zGNRs) structures \cite{Yang2D,Jansch-2017,Schwab,Vo,Yang-2016,narita,bottomup,dossel,Schwab,fogel,dossel,naritaivan,Gao,Vo2,Tan,Wang-2018,
keerthi,Huangvai,Gratz-2018,Li-2018,Morin-2018,Huang-2018,Ai-2018,Koga-2018,Cortizo-Lacalle-2017,Wu-2017,Perkins-2017,Mehd-2017,chen-2017,rogers-2017,li-2016,ma-2015,byun-2015,huang-2015}.\\
\indent Figure \ref{fig:GNR_pattern} gives an overview of the GNRs structures typically produced by solution-phase synthesis. In addition to aGNRs and zGNRs, the following GNRs can be defined: cove-shaped GNRs (cGNRs) and chevron-shaped GNRs. A cGNR is a zGNR wherein a benzo ring periodically decorates the zigzag edge [Fig. \ref{fig:GNR_pattern}(a)]. According to the nomenclature in Ref. \onlinecite{osella}, we label those GNRs as $n$-cGNRs, where $n$ indicates the width of the zGNR core calculated as the number of zigzag lines across the width. On the other hand, chevron-like GNRs can be composed by two segments of regular armchair GNRs with alternating widths connected at a specific angle; these GNRs are called $m/m'$-aGNR \cite{ivan,osella}, Figure \ref{fig:GNR_pattern}(b). The parameters $m$ and $m'$ represent the number of armchair chains that define a specific chevron-like GNR.\\
\indent Raman spectroscopy is the most used non-destructive technique for the characterization of sp$^2$-bonded carbon nanostructures \cite{jorio,cinzia}. It is therefore relevant to investigate whether (and how) the GNRs Raman spectrum provides clear fingerprints of the details of the ribbon atomic structure, which are known to crucially affect their overall electronic and optical properties. A seminal work from our group \cite{ivan} has already shown that the Raman spectrum of cGNR, in particular the low-energy (i.e. acoustic) region, is very sensitive to the corresponding atomic structure, in particular to their width, and departures from simple models obtained for aGNR and zGNR were observed. Some Raman spectra of the 9/15 a-GNRs were also presented in Ref. \onlinecite{Huangvai}, but a detailed multi-wavelength analysis was not reported. Here we expand and generalize our analysis by investigating more than 12 GNRs with different structures, in order to rationalize the D- and G-peak Raman signatures and their dispersion in terms of ribbon properties, such as edge pattern, length, number, position and type of functional groups. We remark that such study is not a mere zoology, but it is mandatory to unravel the origin of the main Raman peak dispersions, as we will show in this work.

Herein we present a detailed multi-wavelength Raman analysis of fifteen GNRs with different structures, including cove-type and chevron-like GNRs. In the case of cGNRs we investigated ribbons with the same width, but functionalized with different groups, also exploring the case of GNRs with low functionalization efficiency.
In the case of chevron-like GNRs, we investigate the effect of the GNR length and the type of functional group. The overall features of the Raman spectra are then compared for all ribbons. We find that all spectra show a relatively intense D peak with a characteristic energy dispersion depending on the precise structure of the GNR -- no matter which functional group is used, provided it is larger than a single atom simply passivating C edge atoms. The dispersion is also independent of the length of the ribbons in the range here investigated ($5 - 110$ nm). We also observed G peak dispersion, although only in few cases, the value of the dispersion is larger than that observed for higly defective graphene. Moreover, the functionalization is observed to strongly affect the low energy region, in agreement with the preliminary results reported in Ref. \onlinecite{ivan} for limited cases. Notably, GNRs with sizeable functional groups at the edges usually do not exhibit a clear Radial Breathing-like Mode, in contrast to the case of carbon nanotubes and of GNRs passivated with single atoms.\\
\indent This paper is organized as follows: \mbox{section \ref{sec:bkg}} gives a background on Raman spectroscopy of GNRs; \mbox{section \ref{sec:methods}} describes the specific systems under investigation and provides details on samples preparation, experimental setup and Raman Spectroscopy measurements, as well as computational approaches. \mbox{Section \ref{sec:results}} shows and discusses the Raman multi-wavelength results obtained for different types of GNRs.

\begin{figure*}[!t]
\centering
\includegraphics[width=0.95\textwidth]{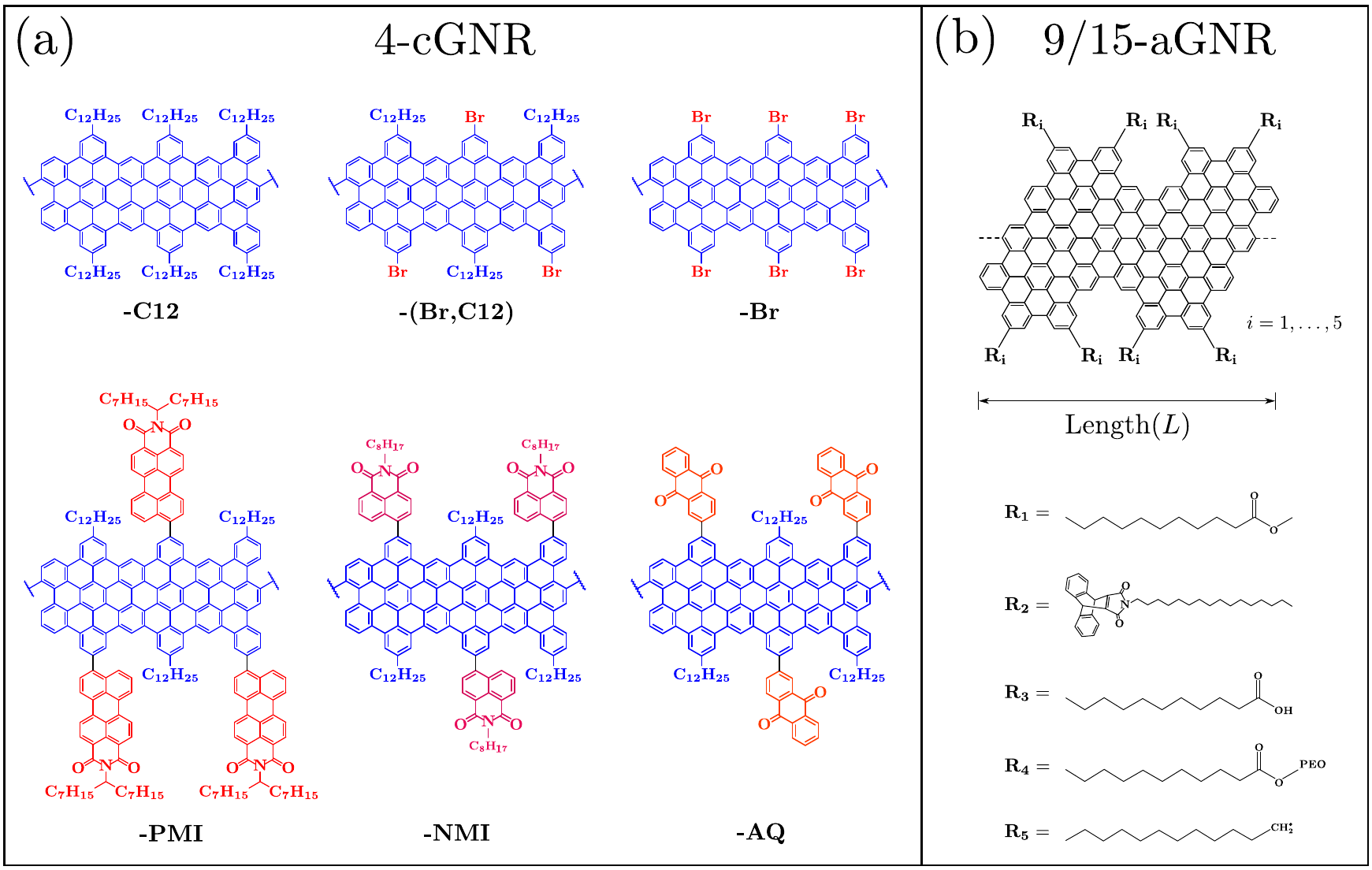}
\caption{Chemical structures of the GNRs studied in this work: (a) cove-shape and (b) chevron-type.}\label{fig:allGNRstructure}
\end{figure*}

\section{\label{sec:bkg}Background}

The Raman spectrum of graphene is composed by two main features, the G and the 2D peaks, which lay at around \mbox{$1580$ \cm-1} and $2800$ \cm-1, respectively \cite{bggraphene}. The G peak is a first-order Raman mode, which arises from the stretching of the carbon atoms and is common to all sp$^2$ carbon systems \cite{bggraphene}; it corresponds to the optical $\text{E}_{\text{2g}}$ phonon at the Brillouin zone center \cite{jorio,Ferrarinano}. The 2D peak is a second-order overtone mode, corresponding to transverse optical (TO) modes near the K point and it is associated with the breathing modes of six-atom rings \cite{bggraphene}. In the presence of defects, these modes can be activated by an inter-valley double resonance process, giving rise to the so-called D peak~\cite{Tuinstra}.\\
\indent The Raman spectrum of GNRs differs from that of graphene and it may considerably change depending on the method used to produce the ribbons \cite{graphita}. In general, the Raman spectrum of GNRs shows the following features:
\begin{enumerate}[label=(\roman*)]
\item the G peak is upshifted and broader, compared to the G peak of (undoped) graphene \cite{ivan};
\item similarly to the case of polyaromatic hydrocarbons (PAHs) \cite{Maghsoumi}, the vibrations corresponding to the D peak can be Raman active and do not necessarily require defects to be active \cite{Tuinstra,Dpeakdisp}; the D peak is typically ``structured'', i.e. consisting of several components, where one of them is more prominent than the others \cite{ivan};
\item in cGNRs the D peak has been observed to change position with the excitation wavelength \cite{ivan}, i.e. the D peak is dispersive. The origin of this characteristic dispersion -- different from that of graphene \cite{Dpeakdisp} -- is currently unknown, and will be discussed in this work;
\item ultra-narrow GNRs (width $\ll 10$ nm) show a characteristic Raman mode at low energy, associated to the vibrational mode where all the atoms of the ribbon move in-plane along the ribbon width direction. This is called Radial Breathing-like Mode (RBLM) \cite{jia,Gillen}, and it was first observed in armchair GNRs produced by surface-assisted methods \cite{cai}. For cGNRs, this mode was demonstrated to be very sensitive not only to the lateral size, but also to any modification of the edge structure\cite{ivan}. A more systematic study is however still missing.
\end{enumerate}

\begin{figure*}[t!]
\centering
\includegraphics[width=0.99\textwidth]{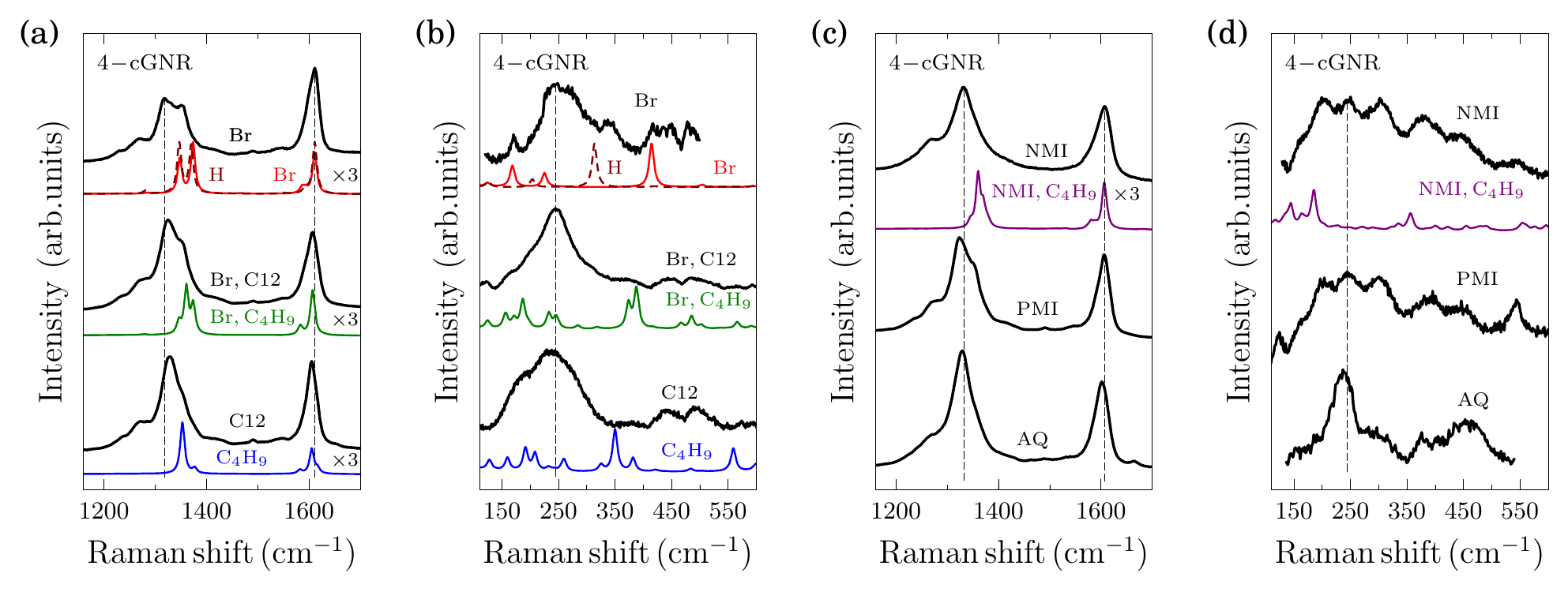}
\caption{(a,c) Optical and (b,d) acoustic region of the Raman spectra for 4-cGNRs with different edge functionalizations. Black (color) lines represent measured (calculated) spectra. In the simulated spectra, the G peak intensity has been rescaled ($\times 3$), compared to the D peak, for clarity.}
\label{fig:all4cGNRspectra}
\end{figure*}

\section{\label{sec:methods}Methods and System Description}

\subsection{\label{sec:systems} Systems under Investigation}

In this work we analyze two classes of GNRs, summarized in Figure \ref{fig:allGNRstructure}.
The first class contains $n$-cGNRs with the same width ($n=4$), but with different edge functionalization [Fig. \ref{fig:allGNRstructure}(a)], that is, the outer benzo rings are functionalized with: dodecyl chains (-C12), bromine (-Br), perylene monoimide (-PMI), naphthalene monoimide (-NMI) and anthraquinone (-AQ) units. Note that, as displayed in  Figure \ref{fig:allGNRstructure}(a), there is one functional group and one dodecyl chain per repeating unit, but the substitution pattern is random.
The second class of systems is represented by chevron-like GNRs. Within this family, we specifically consider the \mbox{9/15-aGNR}, Figure \ref{fig:allGNRstructure}(b). The \mbox{9/15-aGNRs} have been functionalized with five different functional groups: methyl undecanoate (C$_{12}$H$_{24}$O$_2$, $\text{R}_1$), anthracenyl units and N-n-hexadecyl maleimide (AHM, $\text{R}_2$), undecanoic acid (C$_{11}$H$_{22}$O$_2$, $\text{R}_3$), undecanoate grafted with poly(ethylene oxide) chain (C$_{11}$H$_{20}$O$_2$(PEO), $\text{R}_4$) and dodecanyl (C$_{12}$H$_{25}$, $\text{R}_5$) groups.
In the case of $\text{R}_1$ and $\text{R}_2$, three GNRs with different length ($15$ nm, $30$ nm and $110$ nm; $5$ nm, $11$ nm and $58$ nm, respectively) were produced.

\subsection{\label{sec:synthesis} GNR synthesis}

4-cGNR-C12 was synthesized through an AB-type Diels-Alder polymerization of a tetraphenylcyclopentadienone-based monomer to afford polyphenylene precursors, and subsequent intramolecular oxidative cyclodehydrogenation, namely ``graphitization'', using FeCl$_3$ to obtain GNR-C12 \cite{narita}. 4-cGNR-Br was then prepared by adapting the synthetic method of GNR-C12, using a bromo-functionalized monomer precursor \cite{asia}. The dye-functionalized GNRs, i.e., 4-cGNR-PMI, -NMI and -AQ, were synthesized by adapting the pre-functionalization protocol previously reported for hexa-peri-hexabenzocoronene derivatives \cite{keerthi}. Finally, 9/15-aGNRs were synthesized as described in Refs. \onlinecite{asia,huang,Huangvai}.

\subsection{\label{sec:raman_method} Raman Spectroscopy Measurements}

The GNRs samples described above were measured in powder form by using two Raman spectrometers: the XploRA PLUS by Horiba and a Renishaw InVia. Both instruments are equipped with several excitation lines in the visible and near-IR range. The laser spot was about $0.5\,\mu\text{m}$  and was focused on the sample with a \mbox{100$\!\times$} objective. It has been shown \cite{ivan} that ultra-narrow GNRs are very sensitive to the laser power, so this was kept below $550 \,\mu \text{W}$ to avoid damage and ensure reproducible measurements.

The low-energy modes, D and G peaks have been fitted with Lorentzian lineshape. In the case of the D peak, because of its ``structured'' nature, the most prominent (i.e. intense) peak has been considered as the ``D peak'' in the data analysis (see Sec. \ref{sec:results}). The same protocol is applied when fitting the low-energy modes in case of multi-Lorentzian fitting. The intensity is calculated as the height of the Raman peak. The spectra of \mbox{4-cGNR-C12} are taken from Ref. \onlinecite{naritaivan}. All Raman spectra and representative fits are included in the Supplementary Information \cite{Supp_info}.

\subsection{\label{sec:theo_method} Computational Details}

All of the computational results presented in this work were obtained by means of {\it ab initio} density-functional-theory (DFT) based simulations, as implemented in the Quantum ESPRESSO package \cite{Giannozzi1,Giannozzi2}. In particular, the vibrational properties were computed, starting from the optimized geometries of selected GNRs, within the framework of Density-Functional Perturbation Theory (DFPT) \cite{baroni}; Raman intensities were derived using the second-order response method in \mbox{Ref. \onlinecite{maurilazzeri}}, within the Placzek approximation (i.e. non-resonant condition), by using the non-polarized formula in Ref. \onlinecite{Porezag1996}.
The exchange correlation potential was evaluated through the local density approximation (LDA) with the Perdew-Zunger (PZ) parameterization, using Von Barth-Car (VBC) norm-conserving pseudopotentials. The plane-wave energy cut-off was set to 90 Ry. Due to periodic boundary conditions, a vacuum region of at least 12 \AA\, was set in the non-periodic (finite) directions in order to prevent spurious interactions with system replicas. The atomic positions were fully relaxed until the maximum atomic force was less than $5 \times 10^{-4}$  a.u. Phonon frequencies and Raman tensor were calculated by performing the numerical integration of the Brillouin zone (BZ) over a {\bf k}-mesh of dimensions $8\times1\times1$. The calculated phonon frequencies were scaled by a constant factor in order to partially compensate the underestimation of the bond lengths in LDA calculations \cite{Gillen,Gillen2,Rosenkranz} and thus better compare with Raman measurements. The scaling factors were found for each GNR by fixing the simulated G peak frequency to the corresponding experimental value.
\begin{figure}[!t]
\centering
\includegraphics[width=0.89\columnwidth]{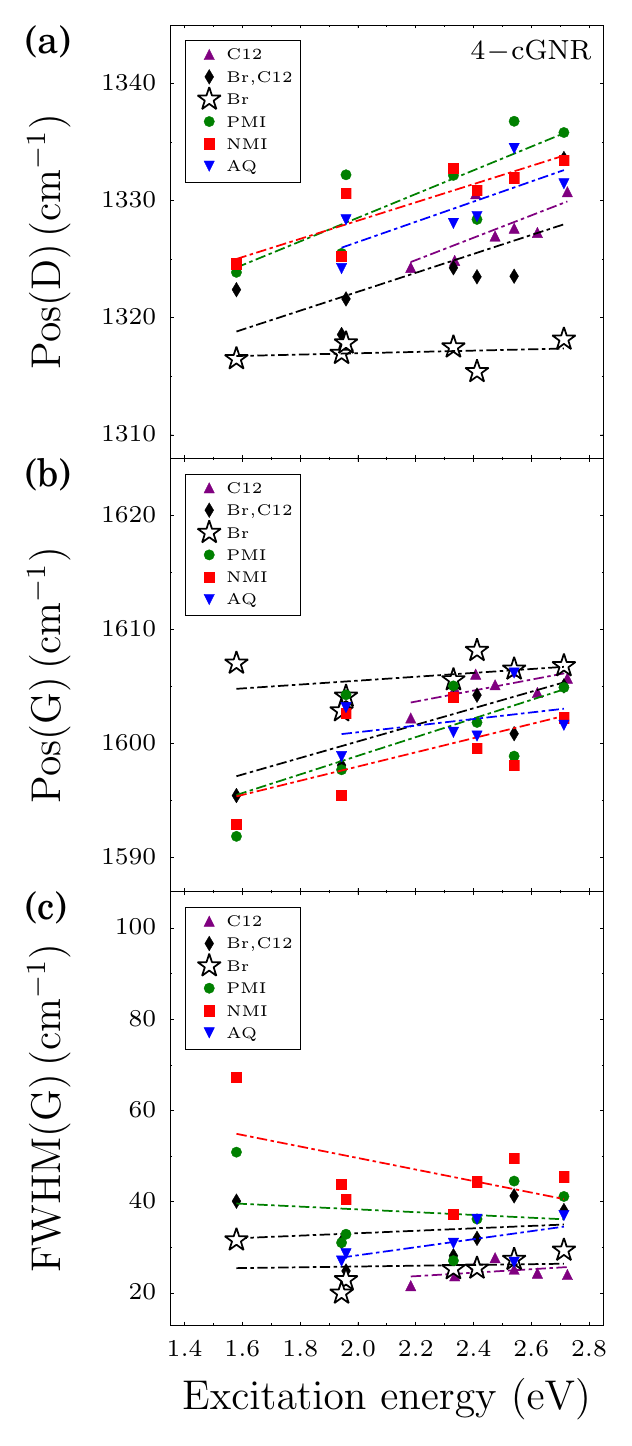}
\caption{Evolution of the position of the (a) D peak, (b) G peak, and (c) FWHM of G peak versus excitation energy for 4-cGNRs. The dash-dotted lines are the linear fits of the data.}\label{fig:alldisp4cGNR}
\end{figure}

Since DFPT simulations of these systems are computationally very demanding, we limited our computational study to prototypical cases. In particular, we investigated the cove-shaped GNR family: fully Br-passivated 4-cGNRs; 4-cGNR half-passivated with Br and alkyl chains as prototype for the -(Br,C12) functionalization; 4-cGNR half-substituted with -NMI groups and alkyl chains as representative for the bulky -NMI functionalization. The simulated spectra for the 4-cGNR - both fully H-passivated and fully functionalized with alkyl chains - are taken from Ref.~\cite{ivan}. Note that butyl chains are used in place of dodecyl and 4-decylhexadecyl groups. A detailed analysis of the effect of the chain length and conformation onto the Raman spectra is reported in Ref.~\cite{ivan}.

\section{\label{sec:results}Results and Discussion}

\subsection{Main Raman features: D, G and low-energy peaks measured at fixed wavelength}

\subsubsection{\label{coveGNRs}Cove-like GNRs}
\begin{figure}[!t]
\centering
\includegraphics[width=0.9\columnwidth]{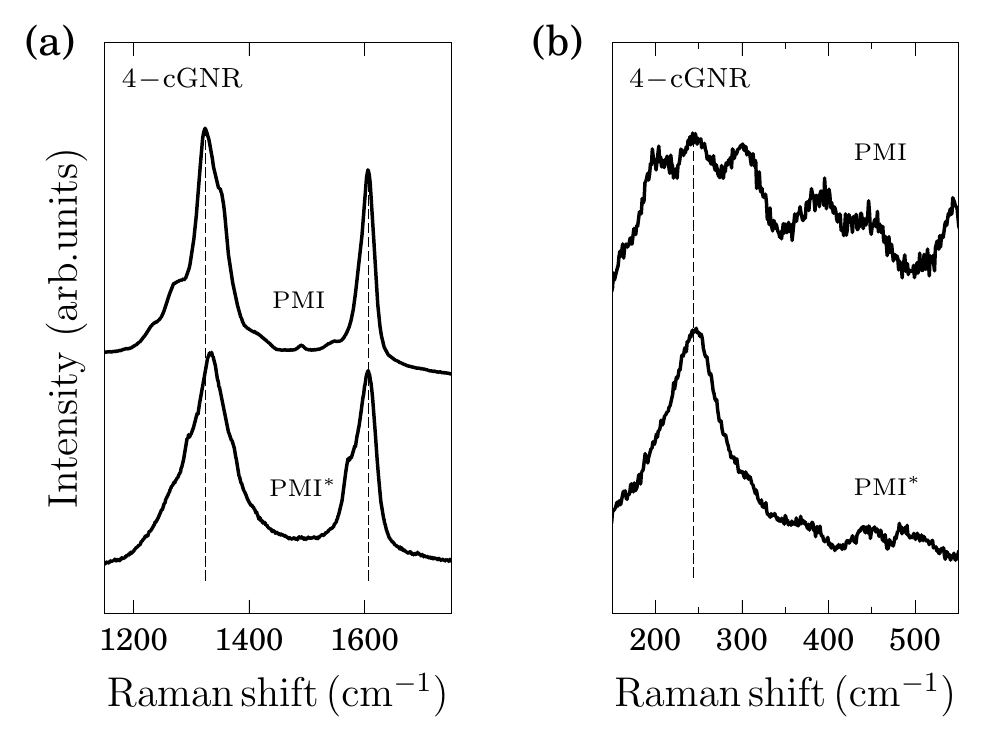}\\
\includegraphics[width=0.89\columnwidth]{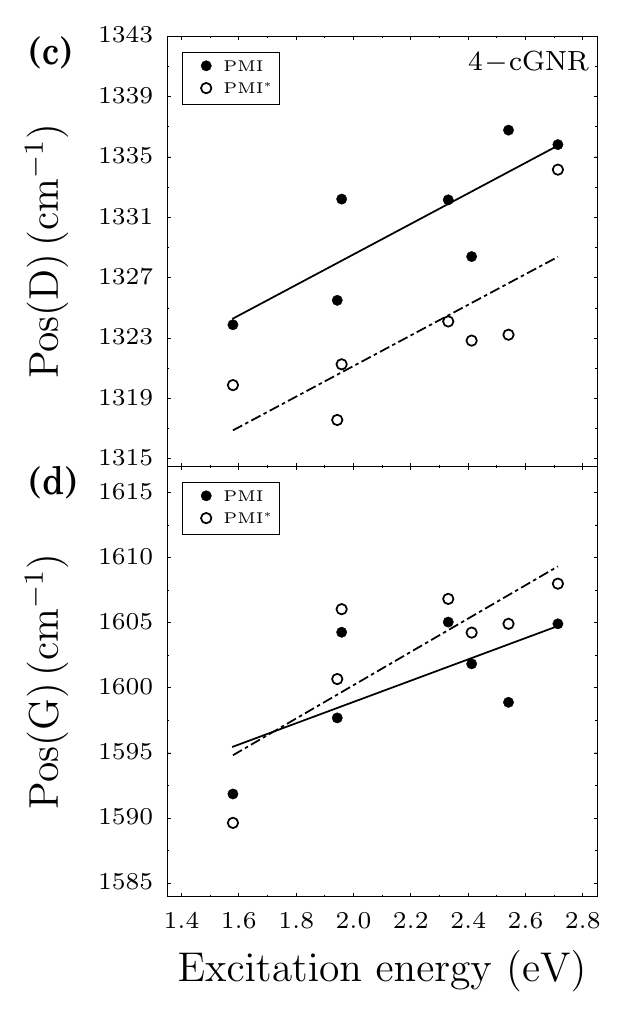}
\caption{(a) Optical and (b) acoustic region of the Raman spectra for the 4-cGNR-PMI$^*$, as compared to the -PMI case. Evolution of the position of the (c) D (d) G peaks versus excitation energy for -PMI and -PMI$^*$. The solid lines show the linear fits for the data obtained for -PMI, and the dash-dotted lines for $\text{-PMI}^*$.}\label{fig:spectra-dispPMI}
\end{figure}
\begin{figure}[!t]
  \begin{center}
    \includegraphics[width=0.9\columnwidth]{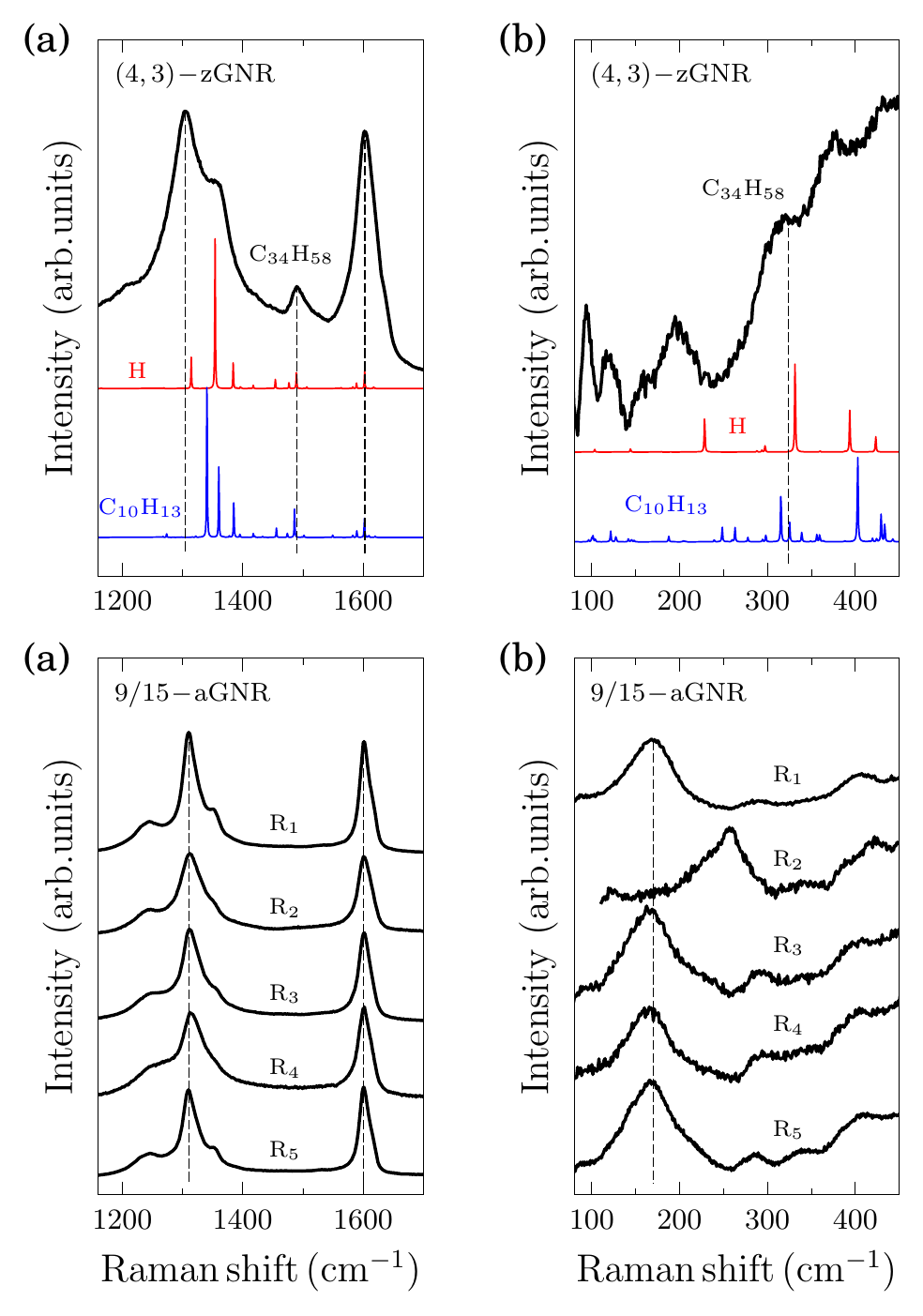}
  \end{center}
  \caption{(a) Optical and (b) acoustic region of the Raman spectra for 9/15-aGNRs with different edge functionalizations. The spectra are taken at 638 nm.}
\label{fig:allspectrachGNR}
\end{figure}
Figure \ref{fig:all4cGNRspectra} displays the Raman spectra of the 4-cGNRs shown in Fig. \ref{fig:allGNRstructure}(a). For the sake of comparison, all Raman spectra discussed hereafter are measured at 2.33 eV. Let us start our investigation from the fully brominated system (-Br), which is the closer system to the ideal fully hydrogenated case usually taken as a reference. The experimental Raman spectrum of 4-cGNR-Br [Fig. \ref{fig:allGNRstructure}(a)] shows a G peak at around 1605 \cm-1, upshifted and broader with respect to graphene. The analysis of {\it ab initio} simulations shows that the G peak mainly consists of the transverse-optical component (0-TO) in addition to weaker contributions at lower wavenumbers due to higher LO and TO harmonics (Fig. S12, Supplemental Material \cite{Supp_info}), similar to our previous results \cite{ivan} for H-passivated 4-cGNR, here reported for comparison. In addition to the G peak, the Raman spectrum shows a typical, structured D peak at about $1315 - 1335$ \cm-1, with one (or more) weaker contributions at lower and higher wavenumbers. The most prominent feature consists of a double-peak structure with comparable intensity, which is relatively well reproduced by the simulations (despite an upshift of about 30 \cm-1 due to the functional choice). Our simulations indicate that the double-peak structure originates from the two components (i.e., longitudinal and transverse) of the six-atom ring breathing, which are no longer degenerate, in contrast to the case of graphene. This is a clear consequence of confinement, and in particular of the ultra-narrow widths achieved by bottom-up techniques.\\
\indent In the presence of edge functionalization with -C12, we observe that the G peak position and FWHM(G) [Fig. \ref{fig:alldisp4cGNR}(c)] remain nearly unaltered. More significative changes are observed for the D peak: the clear splitting observed in the fully brominated system is less visible when the 4-cGNR is functionalized: small changes are observed as soon as C12 groups are introduced in addition to Br [see -(Br,C12) in Fig. \ref{fig:allGNRstructure}(a)], and then the splitting almost disappears in the -C12 system. This trend is in agreement with the simulations, which indicate that the D peak of the -C12 ribbon is dominated by an intense band ascribed to the symmetric breathing mode of alternated hexagonal rings in the ribbon core (Figs. S13-14, Supplemental Material \cite{Supp_info}). The effect of the chains is thus to partially wash out the asymmetry induced by the extreme confinement.\\
\indent Finally, when bulky dye groups are added to C12, the frequency of the D peak tends to increase as compared to the brominated case, while it slightly decreases for the G peak. Moreover, the FWHM(G) increases from about 25 to about 40 \cm-1, with the spectrum of 4-cGNR-NMI showing the broader FWHM(G) [Fig. \ref{fig:alldisp4cGNR}(c)]. This can be explained by analyzing the simulations for the NMI case, where additional C-C modes of the dye appear in this range, contributing to the G peak broadening.\\
\indent Moving to the acoustic region, the experimental Raman spectrum of fully brominated 4-cGNR shows a clear and sharp peak at \mbox{170 \cm-1}, and a broader one at \mbox{$\sim$250 \cm-1} [Fig. \ref{fig:all4cGNRspectra}(b), black]. Other peaks are observed at 300-500 \cm-1, although the spectrum is relatively noisy in that region.
\begin{figure}[!t]
\includegraphics[width=0.89\columnwidth]{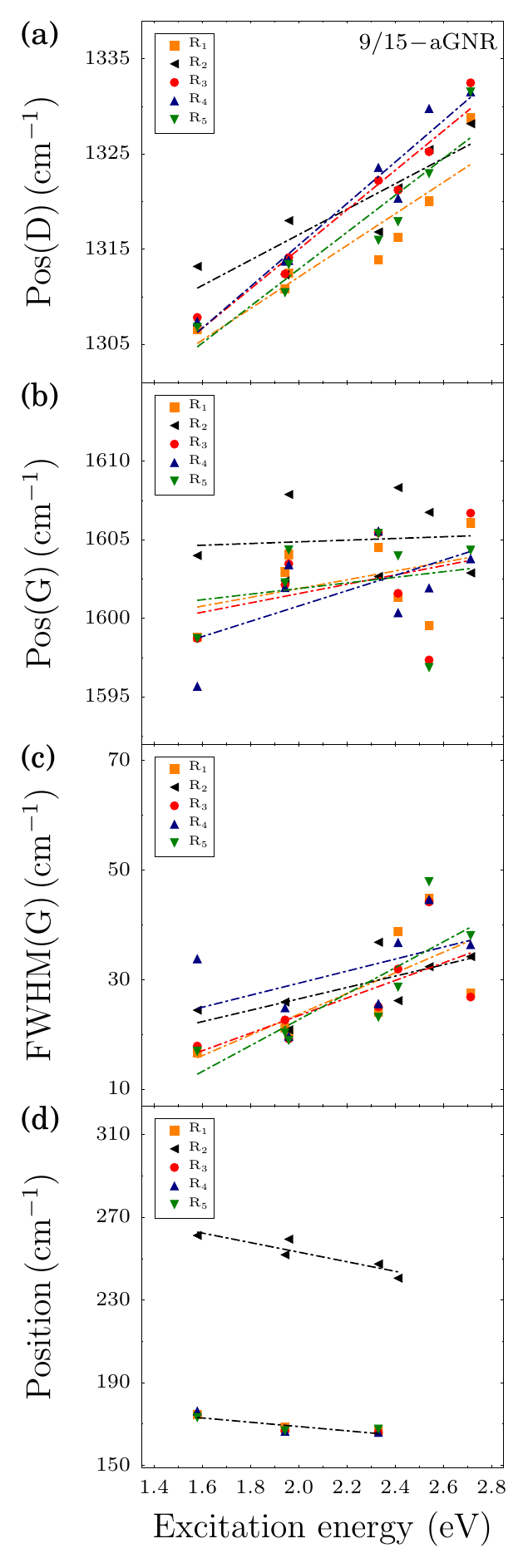}
\caption{Evolution of the position of the position of the (a) D peak, (b) G peak, (c) FWHM of G peak, and (d) low-energy peaks versus excitation energy for 9/15-aGNRs. The dash-dotted lines are the linear fits of the data.}
\label{dispersionschevron}
\vspace{-1cm}
\end{figure}
\noindent These features are in good agreement with those shown by the simulated spectrum, which displays two RBLMs at 168 \cm-1 and 414 \cm-1, where the bromine atoms at the edges vibrate respectively in-phase or out-of-phase with the carbon atoms of the GNR backbone (Fig. S12, Supplemental Material \cite{Supp_info}). This is different from the case of the fully hydrogenated GNR, which shows a single RBLM peak at 314 \cm-1. The prominent feature observed experimentally at \mbox{250 \cm-1}, which is visible also in \mbox{4-cGNRs} partially and fully functionalized with C12, coincides in energy with the simulated LA feature at \mbox{224 \cm-1} for the fully brominated GNR (Fig. S12, Supplemental Material \cite{Supp_info}). However, the broadening and structured nature of the experimental peak advise against any simple attribution.\\
\indent Concerning the partial and full functionalization with docecyl chains [-(Br, C12) and -C12, respectively], the Raman spectra mainly exhibit a structured peak at $\sim 230 - 245$ \cm-1 [Fig.~\ref{fig:all4cGNRspectra}(b)]. As previously evidenced by {\it ab initio} simulations for wider cGNRs~\cite{ivan} functionalized with chains of various lengths, the presence of the chains does not allow to identify a single RLBM, since the GNR low-energy modes couple with the modes of the chain giving rise to several sub-peaks at frequencies that depend on the chain length. Moreover, the edge functionalization relaxes the planar symmetry of the GNRs, allowing the mixing of longitudinal, transverse and normal modes, thus activating modes otherwise forbidden, and overall contributing to the broadening of the main low-energy peak. In this respect, the partial functionalization with dodecyl chain comprises both features, i.e. Br- and C12-related modes.\\
\indent The functionalization with dyes [-NMI, -PMI and -AQ, Fig. \ref{fig:all4cGNRspectra}(d)] gives rise instead to rather different spectra, depending on the functional group. In the case of -AQ, a main peak at 236 \cm-1 can be identified, i.e. at a similar frequency as for -C12 and -(Br,C12), possibly suggesting it having the same origin. Moreover, weaker peaks appear and they could be attributed to coupled modes \cite{ivan}, since AQ is known to display Raman modes in this energy region \cite{keerthi}. On the other hand, the acoustic region of the Raman spectra of -PMI and -NMI GNRs are similar and a clear, distinct peak attributable to a RLBM cannot be identified. We thus simulated the spectrum for the -NMI functionalization as representative of this second type of behaviour to gain further insight. We indeed found a plethora of low-energy modes, among which we identify a RLBM at \mbox{356 \cm-1} (Fig. S15, Supplemental Material \cite{Supp_info}). As found for the alkyl chains, the RLBM of the GNR couples with the modes of the dodecyl and PMI groups, also relaxing the purely transverse character of the mode.\\
\indent The effect of functional groups on the acoustic region of 4-cGNRs was further investigated by studying -PMI GNRs in which the substitution reaction did not succeed completely. This means that the -PMI functionalization was only partial, i.e. where only a few Br atoms of the 4-cGNR-(Br,C12) precursor replaced by PMI molecules. This case is labelled as 4-cGNR-PMI*. The first order Raman spectrum of -PMI* does not show noticeable differences with respect to -PMI [Fig. \ref{fig:spectra-dispPMI}(a)], except for a downshift of $\sim$ 4 \cm-1 in the D peak position [Fig. \ref{fig:spectra-dispPMI}(c)], similar to what observed in the Raman spectrum of the 4cGNR-(Br,C12). Moving to the low energy region, in the case of \mbox{4-cGNR-PMI}, a distinct RBLM peak could not be observed [Fig. \ref{fig:all4cGNRspectra}(b)] in contrast to the case of -PMI*, where a peak at \mbox{$\sim$250 \cm-1} [Fig. \ref{fig:spectra-dispPMI}(b)] is well visible, as in the case of pure \mbox{4-cGNR-(Br,C12)}. This further confirms that the low energy region is very sensitive to the number (and type) of functional groups at the edge.

\subsubsection{\label{chevronGNRs}Chevron-like GNRs}

Figure \ref{fig:allspectrachGNR} reports the first order and the low energy Raman spectra of 9/15-aGNRs, all taken at 1.94 eV. The optical region is very similar to that of the \mbox{4-cGNRs}: the G peak is located at about 1600 \cm-1 [Fig. \ref{dispersionschevron}(b)], and the FWHM(G) is about 30 \cm-1 for \mbox{9/15-aGNR} [Fig. \ref{dispersionschevron}(c)]. Functionalization produces only small changes ($\sim$10 \cm-1) in the position and FWHM(G) of the \mbox{9/15-aGNR}. The D peak is observed at 1320 \cm-1 for 9/15-aGNRs.
For the 9/15-aGNRs, the low-energy region evidences a broad main peak at 165 \cm-1  and a weaker feature at 290 \cm-1 [Fig. \ref{fig:allspectrachGNR}(b)], except for the case the \mbox{R$_2$-functionalized} ribbon (\mbox{257 \cm-1}). This difference can be attributed to the nature of the functional groups, which are all based on long alkyl chains, apart from R$_2$, which is characterised by a bulky group connected to a chain [Fig. \ref{fig:allGNRstructure}(b)]. This bulky group is indeed responsible for the excellent dispersibility of these GNRs \cite{Huangvai}. Although {\it ab initio} calculations are too demanding in the case of the 9/15-aGNR, we can however estimate the frequency of the RLBM from the model in Ref. \onlinecite{jia} for the ideal, H-passivated system. We obtain a value of 230 \cm-1, which compares rather well with the experimental value (169 \cm-1), considering that a downshift is expected for functionalization with alkyl chains, as previously observed for cove-type GNRs. The peak shift to higher wavenumbers in the case of the $\mbox{R}_2$-functionalization could be due to the steric  hindrance of the R$_2$ bulky groups, possibly inducing backbone distortions. Further study is needed to fully confirm this observation.

\subsection{Multiwavelength analysis}

We have up to now discussed the Raman spectra taken at fixed wavelength, we now move to the multi-wavelength analysis, as many features of the Raman spectrum of GNRs strongly change with the excitation energy, as discussed in the Background section. In particular, we explore excitation energies ranging from 1.57 to 2.71 eV, i.e., pre-resonant and resonant conditions considering the optical gap of these GNRs \cite{keerthi}. We here focus on the excitation energy dependence of both D and G peak for all the studied GNRs.\\
\indent The dispersions of the D and G peak of all ribbons are reported in Table \ref{summarydisp}. Note that typically the G peak dispersion is measured on a wide range of energy, ranging from UV to near-IR \cite{ferrari01,Casiraghi-2005}, to get an accurate value. Here the Raman spectra were measured only in the visible range and some spectra could not be measured at certain energies. This is the case of AQ-GNR: we could not measure the spectrum at 785 nm, due to fluorescence. Thus, the fit is less accurate compared to that of the other ribbons. Thus, we report the value in the table for the sake of completeness, but we exclude it from the discussion of the G peak dispersion. Note also that GNRs get easily damaged under UV laser, so UV Raman spectra cannot be measured.

\begin{table}[!t]
\centering
\renewcommand\tabcolsep{0.12cm}
\renewcommand\arraystretch{2}
\begin{tabular}{c|c|c|c}
\hline
\multicolumn{ 2}{c|}{\textbf{GNRs}}  & \multicolumn{ 2}{|c}{\textbf{Dispersions}} \\ \cline{3-4}

\multicolumn{ 2}{c|}{} &   \textbf{D peak }(\cmeV) &     \textbf{G peak }(\cmeV) \\
\hline
\multirow{6}{*}{\rotatebox{90}{\hspace{-1.9cm}\textbf{Cove-type}}} &         \multicolumn{ 3}{|c}{\textbf{\hspace{1.2cm}4-cGNR}} \\ \cline{2-4}

 &  C12          &    9.5        &    4.8         \\ \cline{2-4}

 &  (Br,C12)          &    8.1        &    7.3         \\ \cline{2-4}

 &  Br          &    0.6        &   1.7         \\ \cline{2-4}

 &  NMI          &    7.8        &    6.2         \\ \cline{2-4}

 & AQ         &    8.6        &    2.9$^{\dagger}$         \\ \cline{2-4}

  &  PMI          &    10.1        &    8.2         \\ \cline{2-4}

   &  PMI$^*$          &   10.1        &    12.8         \\ \cline{2-4}
\hline
\multirow{6}{*}{\rotatebox{90}{\hspace{-0.9cm}\textbf{Chevron-type}}} &     \multicolumn{ 3}{|c}{\textbf{\hspace{1.2cm}9/15-aGNRs}} \\ \cline{2-4}

 &  R$_1$          &    $16.6$        &    2.8         \\ \cline{2-4}

 &  R$_2$          &    $13.4$        &    0.5        \\ \cline{2-4}

 &  R$_3$          &    $20.8$        &    2.9        \\ \cline{2-4}

 &  R$_4$          &    $21.8$        &    4.9         \\ \cline{2-4}

  &  R$_5$          &    $19.4$        &     1.8        \\ \cline{2-4}
\hline
\end{tabular}
\caption{Values obtained for the D and G peaks dispersions of cove- and chevron-shaped GNRs. $^{\dagger}$We remark that in the case of AQ, we could not measure the Raman spectrum at 785 nm, thus the fit is less accurate compared to that of the other ribbons. Thus, we report the value in the table for the sake of completeness, but we exclude it from the discussion of the G peak dispersion.}\label{summarydisp}
\end{table}

\subsubsection{D peak dispersion}

Figure \ref{fig:alldisp4cGNR}(a) shows the dependence of the D peak position with the excitation wavelength for the cove-type GNRs. The fit of the data gives a dispersion of \mbox{8-10 \cmeV} for all 4-cGNRs analyzed, except for the fully brominated case that shows no dispersion, i.e. the data variation is within the resolution of the spectrometer, which is about \mbox{2 \cm-1} [Figure \ref{fig:alldisp4cGNR}(a)]. We remind that the D peak is Raman active for these GNRs, and we would expect no dispersion for ideal, unperturbed GNRs, as indeed observed in the case of edge passivation with monoatomic Br and in the case of GNRs produced by surface-assisted approach \cite{barin2019substrate}. Once C$_{12}$H$_{25}$ chains are introduced, the dispersion does not strongly change with the position and type of dye that is used for functionalization, in addition to the chain. In fact, also in the case of the lowest functionalization control (PMI*), where we expect a random distribution of chains, dyes and Br atoms, the dispersion is unaltered. In a previous work we found different types of cGNRs to have D peak dispersion between 10 and 30 \cmeVno, depending on the width, for a fixed functional group \cite{ivan}. This is in agreement with our results, which indicates that the value of the D peak dispersion is determined by the GNR core geometry, as long as C$_{12}$H$_{25}$ chains are introduced at the edges.\\
\begin{figure}[t!]
\includegraphics[width=0.89\columnwidth]{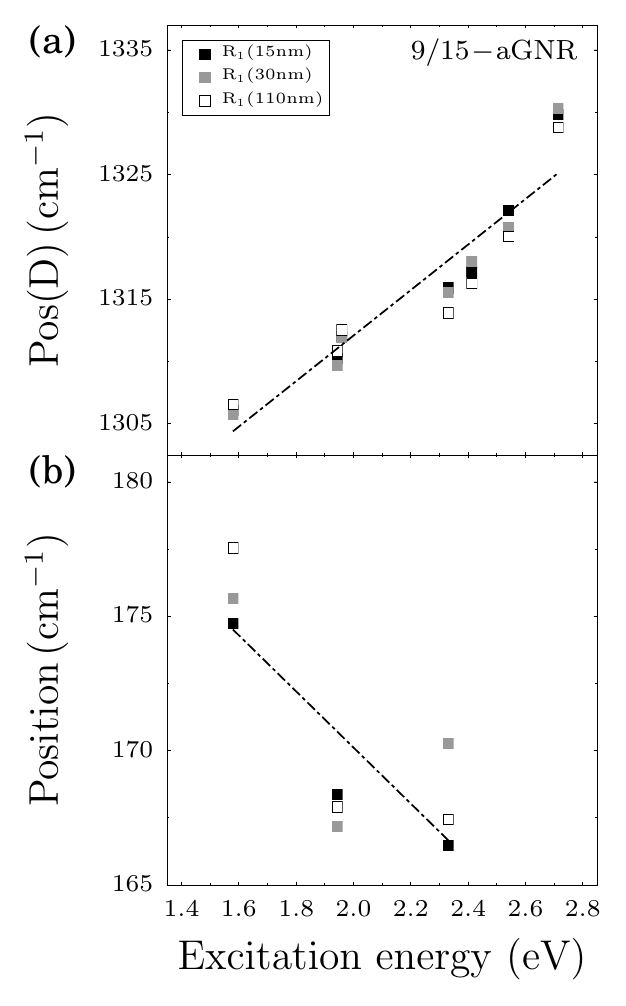}
\caption{Evolution of the position of the (a) D and (b) low energy peaks versus excitation energy for chevron-type \mbox{9/15-aGNRs-R$_1$} with different length ($15$ nm, $30$ nm and $110$ nm). The dash-dotted lines are the linear fits of the data.}
\label{dispersionschevronlength}
\vspace{-0.26cm}
\end{figure}
\begin{figure}[!t]
\includegraphics[width=0.89\columnwidth]{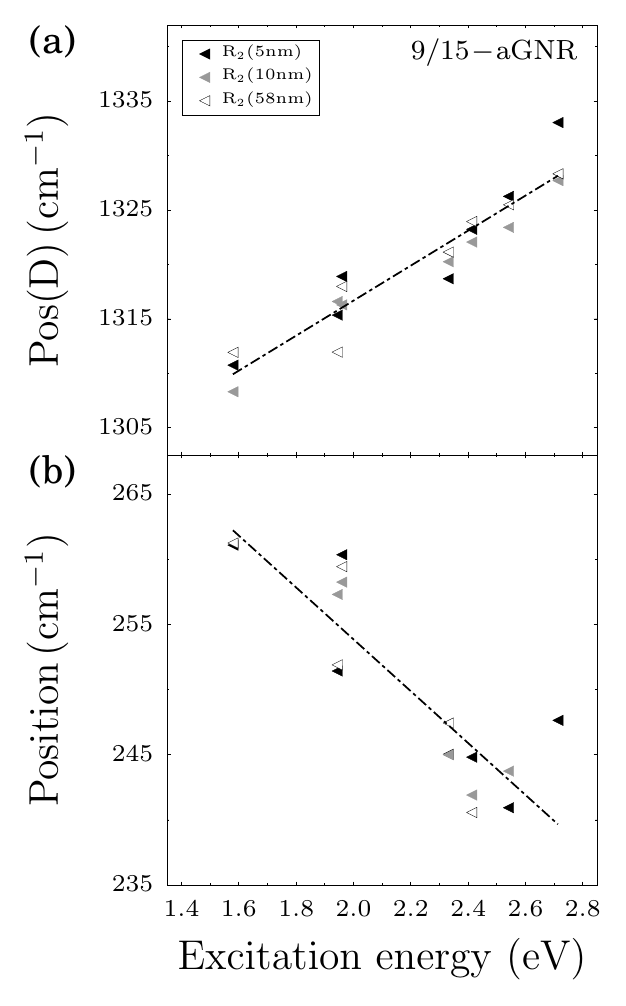}
\caption{Evolution of the position of the (a) D and (b) low energy peaks versus excitation energy for chevron-type \mbox{9/15-aGNRs-R$_2$} with different length ($5$ nm, $11$ nm and $58$ nm). The dash-dotted lines are the linear fits of the data.}
\label{dispersionschevronlengthvai}
\vspace{0.1cm}
\end{figure}
Moving to the chevron-like GNRs, our data show that also the Raman spectra of the \mbox{9/15-aGNRs} exhibit a clear D peak dispersion [Figure \ref{dispersionschevron}(a)] but of higher values (\mbox{17-22 \cmeVno}), as compared to the case of \mbox{4-cGNRs}. Also in this case the dependence on the functional group is negligible, with the exception of the \mbox{9/15-aGNR-R$_2$} case, which shows a significantly smaller D peak dispersion, of about \mbox{13 \cmeVno}. This is likely to be due to the presence of the bulky R$_2$ group at one end of the chain, which could cause steric effects and GNR backbone distortion \cite{Wang-2018}, as also suggested by the strong shift of the main peak in the acoustic region [Figure \ref{fig:allspectrachGNR}(b)].\\
\indent Finally, in order to analyze the effect of the length on the Raman spectra, we compare a series of 9/15-aGNRs with functional groups $\text{R}_1$ [Fig. \ref{dispersionschevronlength}] and $\text{R}_2$ [Fig. \ref{dispersionschevronlengthvai}] with different lengths. To some extent, the ribbon is expected to behave as an elongated aromatic molecule for length shorter that the effective conjugation length, becoming a 1-dimensional nanostructure at longer lengths. We remark that in all cases the length is at least three times larger than the width of the GNRs. In both cases the Raman peaks do not seem to be affected by the length of the ribbon in the range here investigated. In fact the position and the shape of these peaks do not change significantly by increasing the ribbon length from $5$ to \mbox{$110$ nm}, Figs. \ref{dispersionschevronlength} and \ref{dispersionschevronlengthvai} for two representative cases (R$_1$ and R$_2$), which suggests that optical and vibrational properties have already reached the 1d behavior for the shorter GNRs investigated here.\\
\indent To summarize, our analysis has demonstrated that the D peak dispersion is activated anytime the edges are functionalized with long alkyl chains or dye groups, irrespective of the type and position of the functional group, and is characteristic of the GNR core. Moreover, the same dispersion is observed for a wide range of GNR lengths, suggesting that a further confinement along the length is not the main activation mechanism for the D peak dispersion, at least for the length range here considered. All of this points to the key role of the internal degrees of freedom of the functional groups in the activation mechanisms. In fact, the functional chains here considered are flexible and prone to conformational disorder, which can break the translational invariance along the GNRs length and act as a scattering channel.

\subsubsection{G peak dispersion}

We proceed by analyzing the G peak position changes with the excitation energy for both cove-type  [Figs. \ref{fig:spectra-dispPMI}(d) and \ref{fig:alldisp4cGNR}(b)] and chevron-like GNRs [Fig. \ref{dispersionschevron}(b)]. The 4-cGNRs show in all cases a dispersion of \mbox{3-13 \cmeVno}, except for the fully brominated case that shows no dispersion [Fig. \ref{fig:alldisp4cGNR}(b)], i.e. the data variation is within the resolution of the spectrometer, which is about \mbox{2 \cm-1}. In contrast to the D peak dispersion, the G peak dispersion seems to change with the functionalization. The smallest dispersion is seen for \mbox{4-cGNR-C12} and it is about \mbox{3 \cmeVno}. A similar value was observed for the \mbox{8-cGNR-C12} \cite{ivan}, thus indicating that the dispersion value may not be dependent on the core geometry. The G peak dispersion increases as soon as the Br atoms are replaced with the chains and maintain a similar value (6-8 \cmeV) in all cases where the chains are mixed with another type of functional group. The highest G peak dispersion is obtained with the GNR with poor functionalization control, further suggesting the origin of this dispersion to be related to the functional groups. Moving to the chevron-like GNRs, we find that the \mbox{9/15-aGNRs} do not show any appreciable dependence of the G peak with the excitation energy, i.e. the variations in the G peak positions are within the resolution of the spectrometer. This is true irrespective of the functional group and of the GNR length.\\
\begin{figure}[!t]
\centering
\includegraphics[width=0.89\columnwidth]{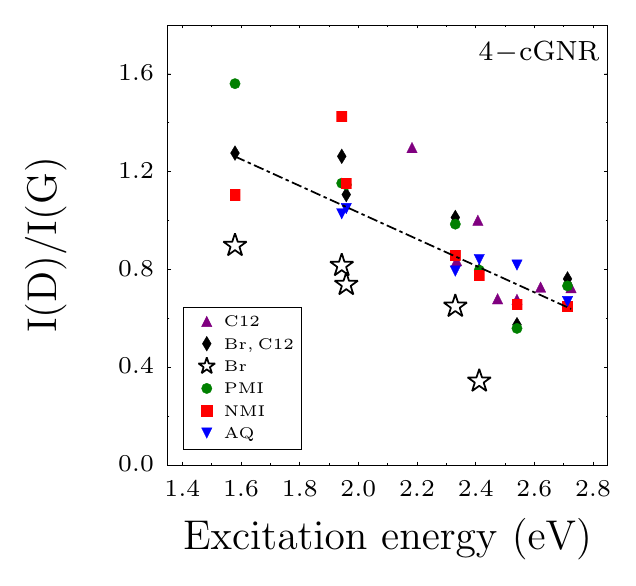}
\caption{I(D)/I(G) data for 4-cGNRs with different edge functionalization, as a function of excitation energy. The dash-dotted line is the linear fit of the data.}\label{fig:alldispIDIG4cGNR}
\vspace{0.3cm}
\end{figure}
\indent In aromatic structures like graphene and PAH, the mode connected to the G peak is not strongly coupled to the $\pi$ electrons, at variance to the ring breathing related to the D peak \cite{castiglioni2004}, thus giving rise to a rather small dispersion. Indeed, a G peak dispersion of only \mbox{6 \cmeV} has been observed in highly defective graphene \cite{cancado}. Most of the GNRs investigated here show small or negligible G peak dispersion, comparable to the G peak dispersion of graphene and compatible with the same activation mechanism suggested for the D peak. However, in a few cases this value is markedly larger (up to 13 \cmeVno), pointing to a different origin for such an excitation energy dependence. While further studies would be needed to clarify this point, we note that in these cases the Br is replaced with chains and dyes. Both Br and the chosen dyes have an electron-withdrawing character, and the dyes present electronic states close to the GNR gap and vibrational states in the same energy region: all of this can impact the intrinsic electronic and optical properties of the ribbon itself \cite{keerthi,cocchi2011optical,cocchi2011designing} and influence the coupling with vibrations in resonant regime. Moreover, the random functionalization pattern can induce localization of the electronic states further influencing the coupling with the vibrations. This might explain why the GNR with no functionalization control (i.e. \mbox{4-cGNR-PMI$^*$}) shows the highest G peak dispersion and why the dispersion of \mbox{4-cGNR-(Br,C12)} is larger than \mbox{4-cGNR-C12}.\\

\begin{figure}[!t]
\centering
\includegraphics[width=0.89\columnwidth]{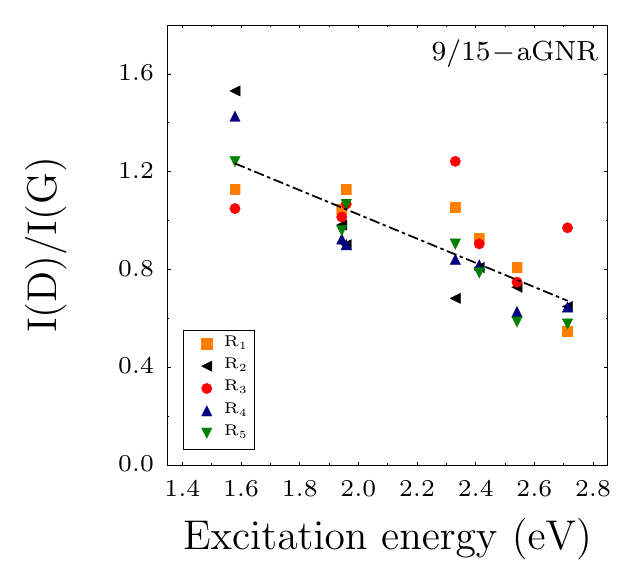}
\caption{I(D)/I(G) for \mbox{9/15-aGNRs} with different edge functionalization, as a function of excitation energy. The dash-dotted line is the linear fit of the data.}\label{fig:alldispIDIGchevrons}
\vspace{0.6cm}
\end{figure}

\subsubsection{I(D)/I(G) intensity ratio}

Concerning the intensity ratio, I(D)/I(G), we find that it depends on the excitation wavelength for both cove-type [Fig. \ref{fig:alldispIDIG4cGNR}] and chevron-like GNRs [Fig. \ref{fig:alldispIDIGchevrons}], as observed also in disordered carbons \cite{ferrari01} and defective graphene \cite{eckmann}. For both GNR families, we do not observe significative changes in the I(D)/I(G) dispersion with the type of functional group. More specifically, both \mbox{4-cGNRs} and \mbox{9/15-aGNRs} exhibit the same I(D)/I(G) dispersion of about $-0.5\text{ eV}^{-1}$.
\vfill{}

\subsubsection{Acoustic region}
Moving to the acoustic region, we focus on the broad peak in the range of \mbox{150-300 \cm-1} for the \mbox{9/15-aGNRs} that might be attributed to the RBLM. Figure \ref{dispersionschevron}(d) shows that the low energy peak is also dispersive: \mbox{$-23 \pm 6$  \cmeV} and \mbox{$-11 \pm 2$  \cmeVno}, in the case of R$_2$ and R$_i$ (i=1,\,3-5), respectively. This is in strong contrast with the results observed for the \mbox{4-cGNRs} \cite{ivan}, where the dispersion was observed to be positive. However, we remark that as the assignments of the peaks is very difficult and we were able to clearly see the peaks only in a very limited range of excitation energies, more studies need to be performed to investigate the origin and dispersions of these peaks.

\section{\label{sec:conclusion}Conclusion}

We presented a detailed multi-wavelength analysis of the Raman spectrum of ultra-narrow cove and chevron-shaped GNRs by focusing on the effect of the excitation energy, edge morphology, type of functional groups, functionalization efficiency and length of the ribbons. We showed that the Raman spectra of such GNRs are characterized by similar features of defective graphene, with typical G and D peaks. In particular, cove-shaped 4-cGNRs show D peak dispersion between 8 and \mbox{10 \cmeV}, with the exception of the fully brominated case. Armchair chevron-shaped GNRs show D peak dispersion ranging between $17$ and \mbox{$22$ \cmeVno}, with the exception of the $\mbox{R}_2$ GNR, possibly due to steric effects. The D peak dispersion does not change with the length of the ribbon, indicating that finite length is not responsible for the dispersion, at least in the ranges investigated here.
By comparison with the Raman spectrum of all Br-passivated 4-cGNR, where the ribbon edges are fully terminated with Br, we observe that the D peak dispersion is activated by the presence of sizeable functional groups at the edge, whose internal degrees of freedom contribute to activate the dispersion. The dispersion is not sensitive to the ``type of defects'', at least for the range of functional groups analyzed in this work and by assuming no steric effects taking place (see R$_2$ functionalization, with 13 \cmeVno). The G peak is also becoming dispersive. While in most of the cases this dispersion can be attributed to the same mechanism responsible for the D peak dispersion, in a few cases significantly larger dispersion values point to an additional effect, possibly associated to a stronger electronic and vibrational coupling with the dye functional groups.\\
\indent Our results show that the low energy region is extremely sensitive to the type of functional group and functionalization efficiency. In general, the acoustic region for functionalized GNRs displays several peaks:  these bands reflect the interplay between RBLM mode, C-H bending modes and normal vibrations of the functional groups at the GNR edges. Moreover, {\it ab initio} simulations showed that bulky groups at the edges relax the purely transverse nature of the RBLM because they induce distortions in the ribbon core giving rise to out-of-plane vibrations.\\
\indent In conclusion, we have shown that Raman spectroscopy is very sensitive to any structural detail of the GNR, making the spectrum more complex to analyze, as compared to that of carbon nanotubes or graphene. On the other side, spectral features, such as D and G dispersions and the peaks at low energy, can provide direct insights on the geometry of the ribbon, by making Raman spectroscopy a very useful tool for the characterization of such \mbox{structures}.

\begin{acknowledgments}
This work is partially supported by the EPSRC (DTP scholarship and CDT Graphene NowNano), by the ERC under the European Union's Horizon 2020 Research and Innovation Programme under grant agreement No. 648417, and by the European Union H2020-EINFRA-2015-1 program under Grant No. 676598, project "MaX - Materials at the eXascale". We acknowledge funding by the National Physical Laboratory in London. Computational resources were provided by the ISCRA program via project ``IsC59\_RAGNO'' on the Marconi machine at CINECA and by the user program of the Center for Functional Nanomaterials, which is a U.S. DOE Office of Science Facility, at Brookhaven National Laboratory under Contract No. DE-SC0012704. Y. Mai appreciates the financial support from National Natural Science Foundation of China (21774076 and 51573091).

\end{acknowledgments}

\end{document}